\documentclass[twocolumn,amsmath,amssymb,longbibliography,aps, prb,floatfix]{revtex4-2}
\pdfoutput=1
\usepackage{hyperref}
\usepackage{url}
\usepackage{color}
\usepackage{acro}
\usepackage{chemmacros}
\usepackage[normalem]{ulem}

\newcommand{\beq}{\begin{eqnarray}}
\newcommand{\eeq}{\end{eqnarray}}

\usepackage{graphicx}

\newcommand\numberthis{\addtocounter{equation}{1}\tag{\theequation}}

\def\blue{\color{blue}}

\def\up{\uparrow}
\def\down{\downarrow}

\usepackage{bbm}

\begin{document}
\title{Microscopic origin of ultranodal superconducting states in spin-1/2 systems}
\author{Yifu Cao$^1$, Chandan Setty$^2$, Laura Fanfarillo$^{1,3}$, Andreas Kreisel$^4$ and P.J. Hirschfeld$^1$}
\affiliation{$^1$Department of Physics, University of Florida, Gainesville, Florida 32603, USA}

\affiliation{$^2$Department of Physics and Astronomy, Rice Center for Quantum Materials, Rice University, Houston, Texas 77005, USA}

\affiliation{$^3$Scuola Internazionale Superiore di Studi Avanzati (SISSA), Via Bonomea 265, 34136 Trieste, Italy}

\affiliation{$^4$Niels Bohr Institute, University of Copenhagen, Jagtvej 155 , DK-2200, Copenhagen, Denmark}

\begin{abstract}
Several unconventional superconductors show indications of zero-energy excitations in the superconducting state consistent with the existence of a so-called Bogoliubov Fermi surface (BFS).  In particular, FeSe isovalently substituted with S seems to acquire a nonzero density of 
states at zero energy at low temperatures as the system goes into the tetragonal phase, consistent with a previously proposed phenomenological 
theory assuming an anisotropic spin singlet pairing gap coexisting with a nonunitary interband triplet component.
Here we search for a microscopic model that can support the coexistence of singlet pairing with other orders, including interband nonunitary triplet pairing with magnetization, and discuss several candidates that indeed stabilize ground states with Bogoliubov Fermi surfaces. We show that with proper choice of the coupling strength of the various orders in our model,  spontaneous breaking of  $C_4$ rotational symmetry is realized at low temperatures. This feature resembles the findings of
recent angle-resolved photoemission experiments in Fe(Se,S) in the tetragonal phase. 
\end{abstract}
\maketitle

\section{Introduction}

 It is expected that strong repulsive Coulomb interactions drive sign changes in the order parameters of unconventional superconductors, typically taking the form of  line or point nodes on the Fermi surface.    There are, however, by now well-known cases where superconductors can develop manifolds of  extended zero-energy excitations   called Bogoliubov Fermi Surfaces (BFS), which have the same dimensionality as the normal state FS.  Interest in superconducting states hosting BFS, referred to as "ultranodal states", has been driven recently by theoretical work in systems with multiple fermionic flavors --  either higher spin or multiple bands -- because it was recognized that such extended nodes are topologically nontrivial\cite{Timm2017, Brydon2017, Brydon2018, Setty2019}.

A multiband, spin-1/2 version of this scenario potentially applicable to Fe-based systems was presented in Refs. \cite{Setty2019,Setty2020}, which included dominant spin singlet pairing, as well as two 
interband triplet terms.  These works showed that in order to generate the ultranodal state, time-reversal symmetry breaking triplet pairings were necessary. 
{The mean field  model} was shown to be characterized by a $\mathbb{Z}_2$ topological invariant corresponding to the sign of the Pfaffian Pf$(H_\mathbf{k})$. Sign changes of { the Pfaffian somewhere in the Brillouin zone} could be induced in the theory by tuning the relative magnitudes of the singlet and triplet order parameters.
Dominant singlet pairing always led to the trivial state {\it unless} the singlet gap was highly anisotropic, in which case the sign change of the Pfaffian could drive a transition to  the ultranodal  state hosting a BFS.

This scenario was applied to the enigmatic Fe(Se,S) material, which is in the normal state well characterized by ARPES measurements and the observation of quantum oscillations revealing changes in correlations and the Fermiology\cite{Reiss2017,Coldea2021}.
It had been shown\cite{Matsuda2018,Hanaguri2018} that this system exhibits simultaneous jumps in the residual density of states $N(0)$ and concomitant abrupt drops in the magnitude of the superconducting gap upon entering the tetragonal phase from the nematic phase at low S substitution.  The BFS were then proposed as a natural explanation for these empirical phenomena.  Entering the ultranodal state was shown to be driven by enhanced intraband singlet anisotropy with increasing S substitution, as observed in experiment. In this situation, the BFS  formed near the momenta where the singlet order parameter fell below the relevant triplet component.
In Ref. \cite{Setty2020}, other signatures of the ultranodal state were %
proposed, which have not yet been confirmed.

Recently, however, an ARPES experiment on Fe(Se,S) in the tetragonal phase \cite{nagashima2022discovery} provided direct evidence of nonzero-area regions of the Fermi  surface exhibiting zero spectral gap in the tetragonal phase,   supporting  the existence of the proposed BFS.  The same experiment also observed a clear $C_2$ symmetry of the spectral gap,  implying that any possible ultranodal state  spontaneously breaks the $C_4$ symmetry normal state.

In Refs. \cite{Setty2019,Setty2020}, the Hamiltonian terms required to produce the BFS were introduced phenomenologically, which enabled neither a deeper understanding of the origin of the pairing interaction, nor a self-consistent framework with which to calculate temperature dependences and relative magnitudes of pairing fields. Therefore the main goal of this work is to construct a microscopic Hamiltonian, which might lead to the observed phenomena with appropriate BFS in the tetragonal phase of Fe(Se,S), including the observed $C_4$ symmetry breaking below $T_c$.

In single band models with spin-singlet superconductivity,  a rather well known type of BFS exists if an external magnetic field is present, namely the Volovik effect,  whereby line nodes are broadened by the Doppler shift of quasiparticles in an orbital field.  Naturally one might expect that spin-driven BFS should also exist in single band models with singlet pairing and itinerant ferromagnetic interactions. Indeed, it was shown in Ref.\cite{karchev2001coexistence} that in 3D the $s$-wave state with coexisting ferromagnetic order has spherical nodal pockets, and is  therefore ultranodal in our language. However, it was subsequently pointed out in Ref. \cite{shen2003breakdown} that this solution with  coexisting orders is not energetically favored compared to the solution with nonmagnetic superconducting order. In Appendix A we show that the same conclusion applies to 2D. In short, this simple one-band model with singlet superconductivity and ferromagnetic interactions does not host stable BFS.

\begin{figure}
\centering
    \includegraphics[width=0.85\linewidth]{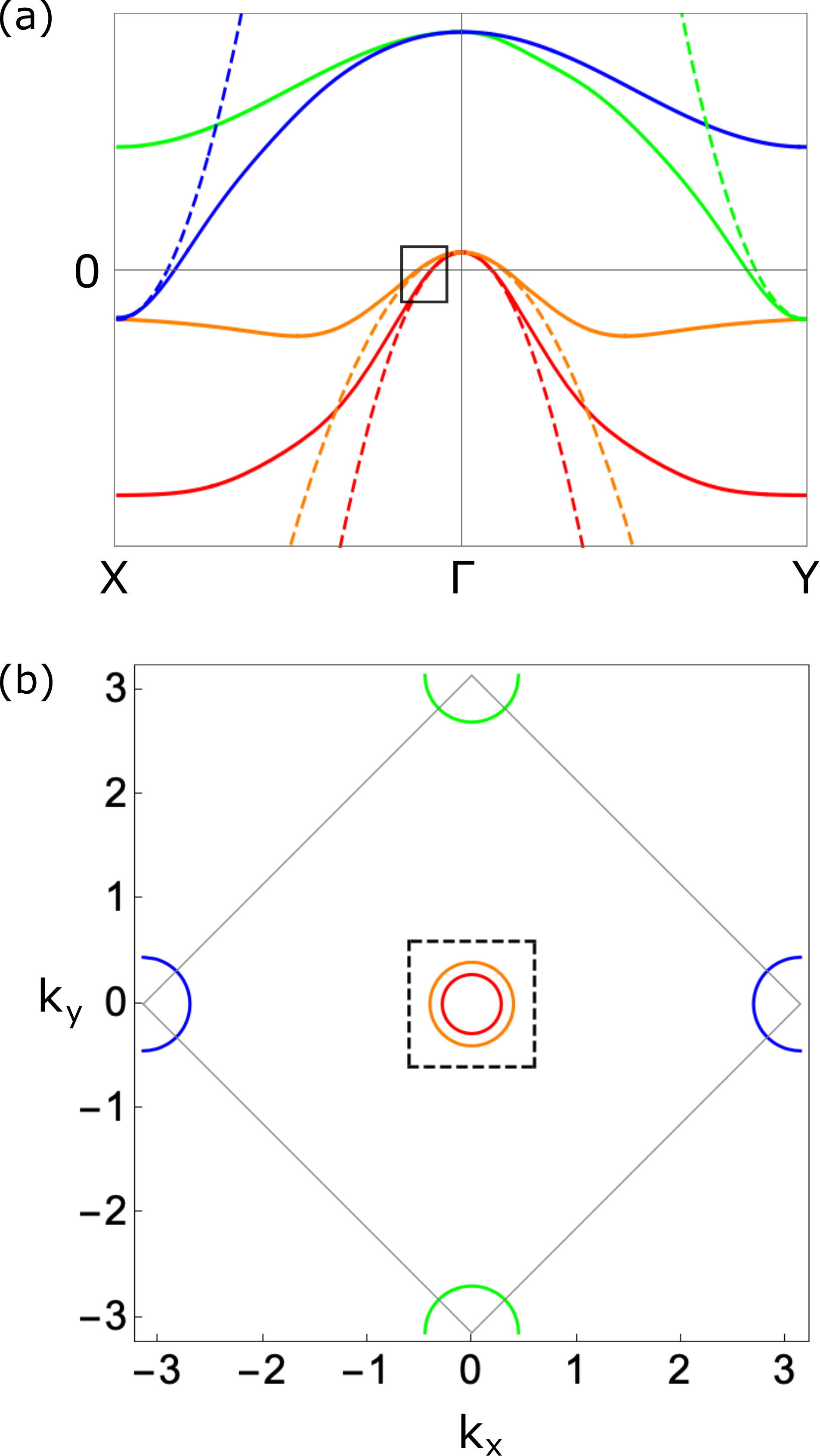}
    \caption {Schematic plots of normal state electronic structure for tetragonal Fe(Se,S) considered in this work. (a) Solid lines are schematic tight-binding bands and the dashed lines are parabolic approximations that are used in our actual calculations. The black square indicates the scope of view for Fig.\ref{Fig1}(a,b) and Fig.\ref{Fig2}(a); (b) Normal state Fermi pockets. The black dashed square corresponds to the scope of view of Fig.\ref{Fig3}(c,d) and Fig.\ref{Fig4}(a) insets.}
    \label{Fig0}
\end{figure}

For the spin-triplet superconducting order parameter considered in our previous proposal \cite{Setty2019,Setty2020}, time reversal symmetry breaking in spin space is required, implying a  nonunitary pairing state, e.g. $|\Delta_{\up\up}|\ne|\Delta_{\down\down}|$.  On the other hand, the phenomena in question are observed in zero external magnetic field, with no ferromagnetic moment above $T_c$.  Thus we search for a {\it spontaneous} condensation of a nonunitary component at or near $T_c$.
First, we briefly review non-unitary triplet superconductivity that is not spontaneous, in single band models.
The theory of non-unitary triplet superconductivity coexisting with itinerant ferromagnetism has been studied extensively,  mainly in the context of single-band models with three dimensional Fermi surfaces relevant to the ferromagnetic superconductors UGe$_2$, URhGe and UCoGe\cite{machida2001phenomenological,nevidomskyy2005coexistence,linder2008coexistence}. In this case, the superconducting $T_c$ for the majority spin is enhanced and the non-unitary state wins energetically over the unitary solution due to the increase in the density of states (DOS) at the Fermi level for the majority spin when shifted by magnetization\cite{linder2007quantum}. In these theories, superconductivity condenses out of a preexisting ferromagnetically ordered state ($T_c<T_{\text{Curie}})$. { The theory of spontaneous non-unitary triplet superconductivity has also been proposed on the Ginzburg-Landau (GL) level by various works \cite{nevidomskyy2020stability,walker2002model,amin2020generalized}. In the presence of a coupling between the non-unitary pairing and the ferromagnetism of the form $\mathbf{m}\cdot (i\mathbf{d}\times\mathbf{d^*})$, spontaneous magnetization and non-unitary triplet state can arise at the same $T_c$ }(here $\bf m$ is the net magnetization and $\bf d$ is the triplet $\bf d$-vector).
It is however not guaranteed that the free energy of the non-unitary state will be lower than that of the unitary triplet state at zero temperature.

For quasi-2D single-band models, the DOS $N(\epsilon)$ is relatively constant near the band edge, and the change in the DOS due to small shifts from the magnetization is negligible. As a result, splitting of subbands of opposite spins does not naturally lead to a splitting of the corresponding equal-spin triplet pair amplitudes as in 3D. Nevertheless, we find that in the case where the Fermi energy is less than the energy cutoffs of superconductivity and magnetism, the $\mathbf{m}\cdot (i\mathbf{d}\times\mathbf{d^*})$ term in the GL expansion is large within weak coupling if the Fermi energy is small.
This should stabilize the non-unitary state near $T_c$. However, we did not find any energetically favorable non-unitary solution that persists down to zero temperature within this single-band model.

In {\it multiband} superconductors with {\it interband} triplet pairing, the splitting due to magnetization can stabilize the nonunitary triplet state much more effectively. These effects lead to possible nonunitary interband triplet ground states even in the absence of preexisting magnetic order.
We discuss this aspect in Sec. II below. In Sec. III, we construct a minimal two-band model that contains interband triplet pairing, fluctuating ferromagnetic order and intraband singlet superconductivity, and show that the self-consistently determined ground state can be an ultranodal state with all  three orders coexisting.  Furthermore, the ground state simultaneously breaks rotational symmetry. In Sec. IV we consider a more realistic four-band model, relevant to the tetragonal (normal state) phase of the Fe(Se,S). In this case the spontaneous $C_4$ symmetry breaking Bogoliubov Fermi surface in our model resembles that reported by the recent ARPES experiment \cite{nagashima2022discovery}. Finally, we discuss the possible interplay between  fluctuating nematic order and the three coexisting orders in our ultranodal solution. We argue that the $C_4$ symmetry breaking ground state within our model can be further stabilized at higher temperatures when taking into consideration the fluctuating nematic order.  This last argument would appear to be relevant to the BFS in Fe(Se,S), which is observed in the tetragonal phase close to the disappearance of the nematic phase.

\begin{figure*}
\centering
    \includegraphics[width=0.9\linewidth]{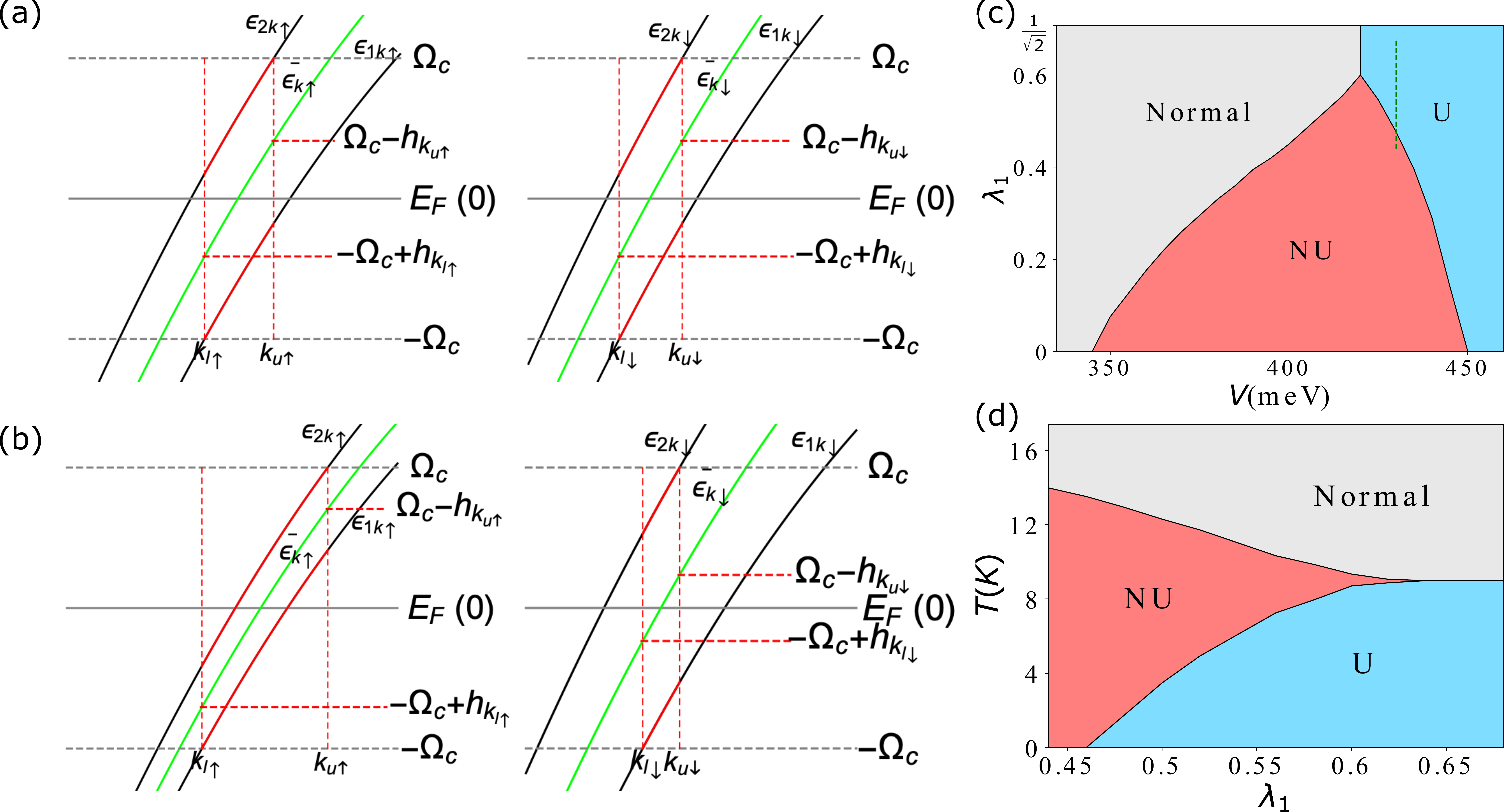}
    \caption{Effects of spontaneous magnetization on interband triplet pairing in a two band model with the same mass $m_1=m_2$. (a,b) Schematic plots of normal state bands shifted by magnetization. Left: the spin up subbands. Right: the spin down subbands.  For each spin, $k_{l\sigma}$ and $k_{u\sigma}$ correspond to the limits of the pairing interaction in momentum space. Case (a): $m_1=m_2$, $\lambda_1=\lambda_2$.  Case (b): 
    $m_1=m_2$, $\lambda_1\neq\lambda_2$.
    Green curves are  averaged dispersion $\bar{\epsilon}_{\mathbf{k}\sigma}$, and $\pm(\Omega_c-h)$ are cutoffs for the energy integral over $\bar{\epsilon}_{\mathbf{k}\sigma}$ (cf Eq.\eqref{gap eq integral in II}). The red color highlights the electrons that take part in the interband pairing.   (c) $\lambda_1-V$ phase diagram at zero temperature of the model Hamiltonian \eqref{H in II}, for $m_1=m_2$.
    Both non-unitary (NU) and unitary (U) triplet phases are chiral p+ip state. Parameters used: $\mu_1={-}10$ meV, $\mu_2={-}15$ meV, $m_1=m_2={-}8$ eV$^{-1}$, $\Delta\theta_c=\frac{\pi}{10}$, $J=3.85$ eV.  Note $J$ is below the Stoner threshold $J_0\approx 4.05$ eV for these parameters. (d) Temperature phase diagram at $V=0.43$ eV. (x-axis corresponding to the dashed line in (c)). All the transition lines are first order { due to the interband nature of the pairing}.}
    \label{Fig1}
\end{figure*}

\section{spontaneous nonunitarity for interband triplet pairing}
We consider a Hamiltonian with two 2D parabolic bands $\epsilon_{i\mathbf{k}}=\frac{\mathbf{k}^2}{2m_i}-\mu_i$ as follow: 
\begin{align*}
    H=&H_0+H_m+H_T\\
    =&\sum_{\substack{\mathbf{k},\sigma\\i=\{1,2\}}}\epsilon_{i\mathbf{k}} c^\dagger_{i\mathbf{k}\sigma}c_{i\mathbf{k}\sigma}\\
    &-\frac{J}{2}\sum_{\substack{\mathbf{k},\mathbf{k'},\sigma,\sigma',i,j}}\Lambda_{ij}(\mathbf{k},\mathbf{k'})\sigma_z\sigma'_zc^\dagger_{i\mathbf{k}\sigma}c^\dagger_{j\mathbf{k'}\sigma'}c_{j\mathbf{k'}\sigma'}c_{i\mathbf{k}\sigma}\\
    &-V\sum_{\mathbf{k},\mathbf{k'},\sigma}\cos(\theta_\mathbf{k}-\theta_\mathbf{k'})c^\dagger_{1\mathbf{k}\sigma}c^\dagger_{2-\mathbf{k}\sigma}c_{2-\mathbf{k'}\sigma}c_{1\mathbf{k'}\sigma}
    \numberthis \label{H in II}
\end{align*}
The two parabolic bands considered here (See Fig. \ref{Fig0}) will correspond to the hole pockets at $\Gamma$ point in a complete four-band model later in Sec. IV. The second term $H_m$ is a magnetic interaction that involves electrons in both bands. $\Lambda_{ij}(\mathbf{k},\mathbf{k'})=\lambda_i\lambda_j\hat{\Lambda}(\mathbf{k},\mathbf{k'})$, where the constant parameters $\lambda_1$ and $\lambda_2\equiv\sqrt{1-\lambda_1^2}$ tune the relative strength of the magnetic interaction on band 1 and 2, and $\hat{\Lambda}(\mathbf{k},\mathbf{k'})$ is a  momentum space cutoff. We use the assumption that $\hat{\Lambda}(\mathbf{k},\mathbf{k'})$ equals $1$ if $|\theta_\mathbf{k}-\theta_\mathbf{k'}|$ is within an angular cutoff $\Delta\theta_c$ and $|\epsilon_{i\mathbf{k}}|, |\epsilon_{j\mathbf{k'}}|$ are both within an energy cutoff, and otherwise $0$. Here $\theta_\mathbf{k}$ is the angle between $\mathbf{k}$ and the $k_x$ axis. Thus the exchange interaction is taken to  affect only electrons with $|\mathbf{k}-\mathbf{k'}|$ within a range in momentum space\cite{Snoke2020book}. In the second term $H_m$ the $\sigma_z^{(')}$ is a shorthand notation for the diagonal elements of the Pauli matrix $\sigma_z$, and it takes value $\pm1$ for spin up/down. The third term $H_T$ is an attractive interband p-wave triplet pairing interaction between equal spins. Note that both the magnetic and pairing interaction in Eq.\eqref{H in II} explicitly break spin rotational symmetry. The fully rotational symmetric interaction will contain $\sigma_z$ terms that is in Eq.\eqref{H in II}, as well as other terms involving $\sigma_{x,y}$. Nevertheless these extra terms would vanish in the mean field approximation even if they were included in Eq.\eqref{H in II} if the magnetization condenses in the z-direction. So we may regard Eq.\eqref{H in II} as representing a system with spin rotational symmetry within a mean field approximation, after condensation of $\bf m$ along $z$. For simplicity, from now on we drop the subscript $z$ and use the notation $\sigma=\pm1$ for spin up/down.

\begin{figure*}
    \centering
    \includegraphics[width=0.95\linewidth]{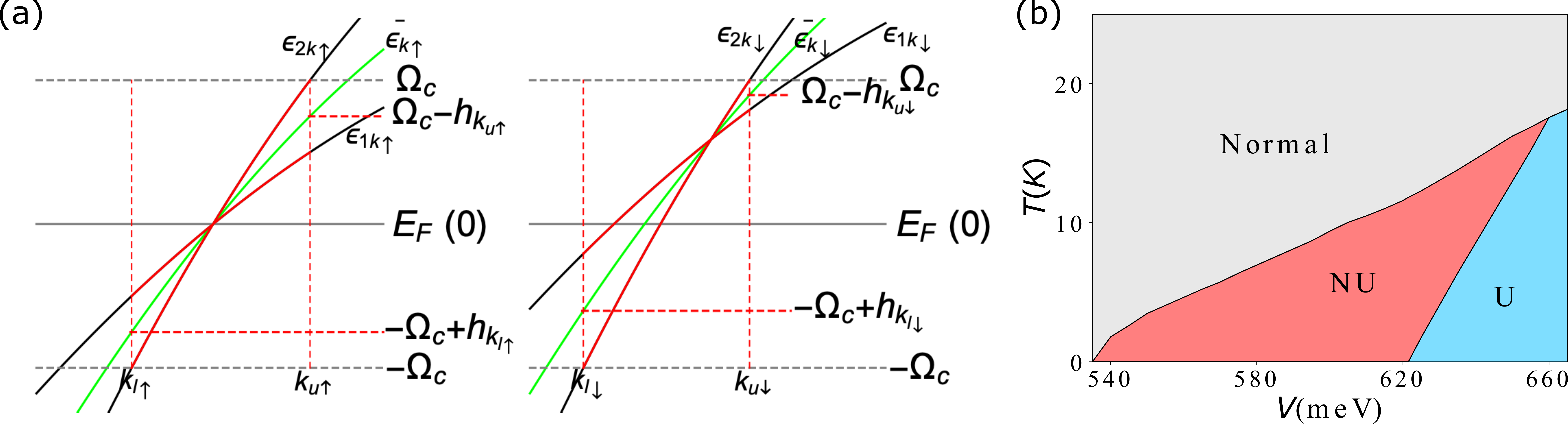}
    \caption{(a) Schematic plots of normal state bands shifted by magnetization, for $m_1\neq m_2$ and $\lambda_1=\lambda_2$. Left: the spin up subbands. Right: the spin down subbands. (b) $T-V$ phase diagram of the model Hamiltonian \eqref{H in II} for $m_1\neq m_2$ and $\lambda_1=\lambda_2$. Parameters used: $\mu_1={-}10$ meV, $\mu_2={-}15$ meV, $m_1={-}8$ eV$^{-1}$, $m_2={-}4$ eV$^{-1}$, $\Delta\theta_c=\frac{\pi}{10}$, $J=5.25$ eV.}
    \label{Fig2}
\end{figure*}

The momentum resolved magnetization on band $i$ is ${\tilde{m}}_{i\mathbf{k}}=\sum_{\sigma}\sigma\langle c^\dagger_{i\mathbf{k}\sigma}c_{i\mathbf{k}\sigma}\rangle$. We define a weighted magnetization $\tilde{m}_\mathbf{k}\equiv\sum_{\mathbf{k'}}\hat{\Lambda}(\mathbf{k},\mathbf{k'})(\lambda_1{\tilde{m}}_{1\mathbf{k'}}+\lambda_2{\tilde{m}}_{2\mathbf{k'}})$. The triplet gaps are $\Delta_{px/py,\sigma\sigma}=V\sum_\mathbf{k} \omega_{px/py}(\theta_\mathbf{k})\langle c_{2\mathbf{-k}\sigma}c_{1\mathbf{k}\sigma}\rangle$, with $\omega_{px}=\cos{\theta_\mathbf{k}}$ and $\omega_{py}=\sin{\theta_\mathbf{k}}$, and the total gap function is $\Delta_{\sigma\sigma}(\theta_\mathbf{k})=\omega_{px}(\theta_\mathbf{k})\Delta_{px,\sigma\sigma}+\omega_{py}(\theta_\mathbf{k})\Delta_{py,\sigma\sigma}$. We also denote the normal state dispersion after shifted by magnetization as $\epsilon_{i\mathbf{k}\sigma}\equiv\epsilon_{i\mathbf{k}} -\sigma J\lambda_i\tilde{m}_\mathbf{k}$.
Thus within mean field approximation the Hamiltonian reads
\begin{align*}
    H=&\sum_{\mathbf{k},\sigma,i}\epsilon_{i\mathbf{k}\sigma}c^\dagger_{i\mathbf{k}\sigma}c_{i\mathbf{k}\sigma}\\
    &-\sum_{\mathbf{k},\sigma}\Delta_{\sigma\sigma}(\theta_\mathbf{k})(c^\dagger_{1\mathbf{k}\sigma}c^\dagger_{2-\mathbf{k}\sigma}+h.c.)\\
    &+\frac{J}{2} \sum_{\mathbf{k},\mathbf{k'},i,j}\Lambda_{ij}(\mathbf{k},\mathbf{k'}){\tilde{m}}_{i\mathbf{k}}{\tilde{m}}_{j\mathbf{k'}}\\
    &+\frac{|\Delta_{px\uparrow\uparrow}|^2+|\Delta_{px\downarrow\downarrow}|^2 + |\Delta_{py\uparrow\uparrow}|^2+|\Delta_{py\downarrow\downarrow}|^2}{V} \numberthis \label{Hmf}
\end{align*}
From above we see that the mean field Hamiltonian is block diagonal in spin space. By diagonalizing the two spin blocks separately, the self-consistency condition can be found as 
\begin{widetext}
\begin{align*}
    &\tilde{m}_\mathbf{k}=\sum_{\mathbf{k'}}\hat{\Lambda}(\mathbf{k},\mathbf{k'})\Big( \frac{\lambda_1-\lambda_2}{2}\big[\delta f (\bar{E}_{\mathbf{k'}\uparrow},h_{\mathbf{k'}\uparrow})-\delta f (\bar{E}_{\mathbf{k'}\downarrow},h_{\mathbf{k'}\downarrow})\big]-\frac{\lambda_1+\lambda_2}{2}\big[\frac{\bar{\epsilon}_{\mathbf{k'}\uparrow}}{\bar{E}_{\mathbf{k'}\uparrow}}th(\bar{E}_{\mathbf{k'}\uparrow}, h_{\mathbf{k'}\uparrow})-\frac{\bar{\epsilon}_{\mathbf{k'}\downarrow}}{\bar{E}_{\mathbf{k'}\downarrow}}th(\bar{E}_{\mathbf{k'}\downarrow}, h_{\mathbf{k'}\downarrow})\big]\Big)
    \numberthis \label{self-consistency1 in II}\\
    &1=V\sum_\mathbf{k}\omega_p^2(\theta_\mathbf{k})\frac{th(\bar{E}_{\mathbf{k}\sigma}, h_{\mathbf{k}\sigma})}{2\bar{E}_{\mathbf{k}\sigma}}
    \numberthis \label{self-consistency2 in II}
\end{align*}
\end{widetext}
where $\delta f (E,h)\equiv f(E+h)-f(E-h)$ and $th(E,h)\equiv1-f(E+h)-f(E-h)$.
Note that Eq.\eqref{self-consistency2 in II} represents four different equations with $p=p_x,p_y$ and $\sigma=\uparrow,\downarrow$. If $h=0$ we have $th(E,0)=\tanh(\frac{\beta E}{2})$.  For fixed $E$ and temperature, $th(E,h)$ is an even function of $h$, and it monotonically decreases as $|h|$ increases. $\bar{\epsilon}_{\mathbf{k}\sigma}$ and $h_{\mathbf{k}\sigma}$ are the average and difference between the two subbands with the same spin, respectively:
\begin{align*}
    \Bar{\epsilon}_{\mathbf{k}\sigma}&=(\epsilon_{1\mathbf{k}\sigma}+\epsilon_{2\mathbf{k}\sigma})/2\\
    &=(\epsilon_{1\mathbf{k}}+\epsilon_{2\mathbf{k}}-\sigma(\lambda_1+\lambda_2)J\Tilde{m}_\mathbf{k})/2
    \numberthis \label{epsilonbar}\\[0.2cm]
    h_{\mathbf{k}\sigma}&=(\epsilon_{1\mathbf{k}\sigma}-\epsilon_{2\mathbf{k}\sigma})/2\\
    &=(\epsilon_{1\mathbf{k}}-\epsilon_{2\mathbf{k}}-\sigma(\lambda_1-\lambda_2)J\Tilde{m}_\mathbf{k})/2 \numberthis \label{h_k,sigma1}
\end{align*}
$E_{\{1,2\}\mathbf{k}\sigma}=\Bar{E}_{\mathbf{k}\sigma} \pm h_{\mathbf{k}\sigma}$ are the Bogoliubov quasiparticle dispersions and  $\Bar{E}_{\mathbf{k}\sigma}=\sqrt{\Bar{\epsilon}^2_{\mathbf{k}\sigma}+|\Delta_{\sigma\sigma}(\theta_\mathbf{k})|^2}$. Eq.\eqref{self-consistency2 in II} can be written in terms of an integral over $\bar{\epsilon}_{\mathbf{k}\sigma}$ and $\theta_{\mathbf{k}}$, as long as $\Tilde{m}_\mathbf{k}$ has only angular dependence on $\mathbf{k}$, which can be deduced from Eq.\eqref{self-consistency1 in II}. To this end, we express $h_{\mathbf{k}\sigma}$ in terms of $\bar{\epsilon}_{\mathbf{k}\sigma}$ and $\tilde{m}_\mathbf{k}=\tilde{m}(\theta_\mathbf{k})$:
\begin{align*}
    h_{\mathbf{k}\sigma}=\frac{1}{m_1+m_2}\big( (m_2-m_1)\bar{\epsilon}_{\mathbf{k}\sigma}
    {\blue -}m_1\mu_1{\blue +}m_2\mu_2\\
    -\sigma(m_1\lambda_1-m_2\lambda_2)J\tilde{m}_\mathbf{k}\big) \numberthis \label{h_k,sigma}
\end{align*}

\begin{figure*}
    \includegraphics[width=\linewidth]{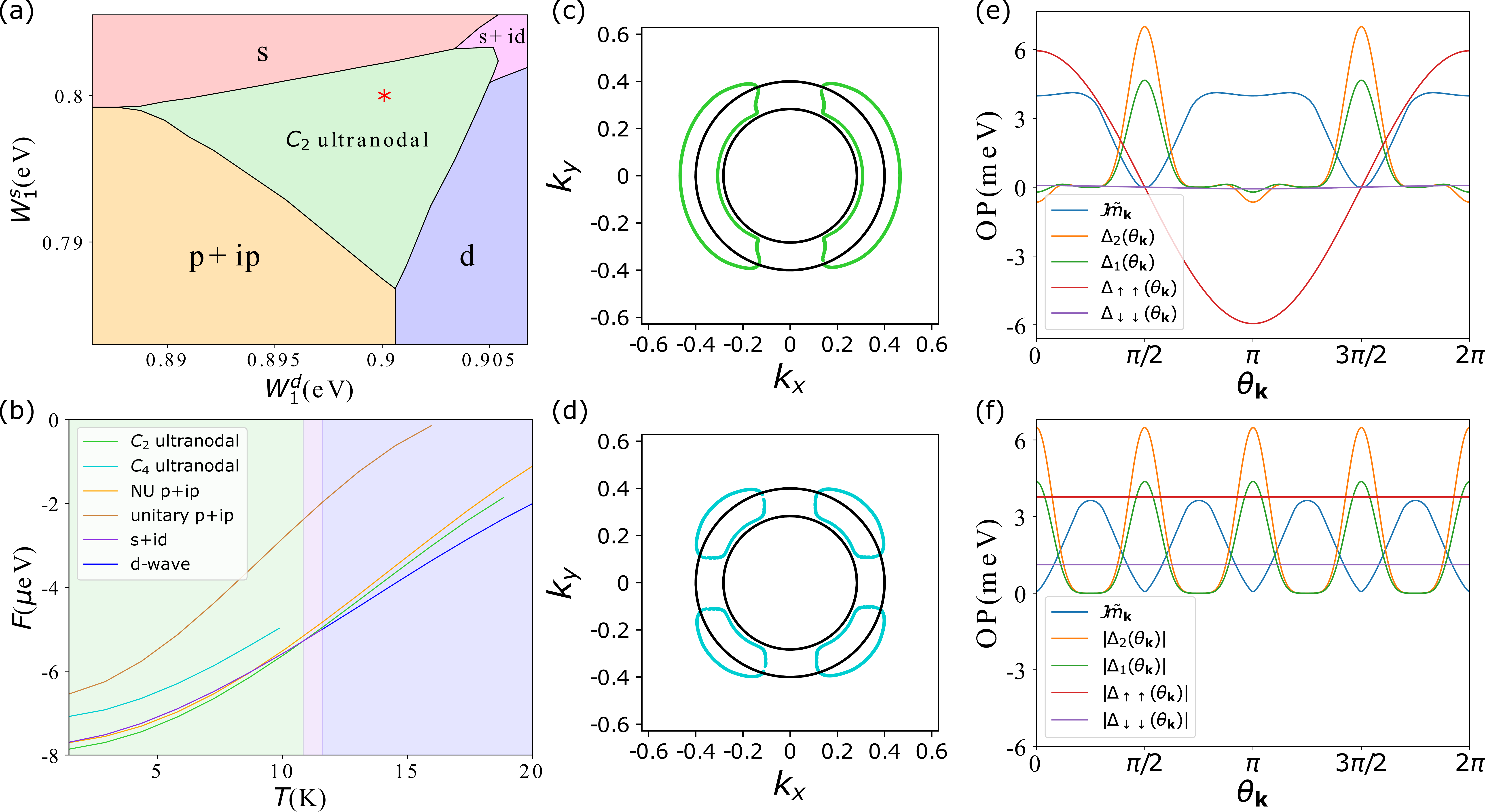}
    \caption{Results for the two band model Hamiltonian \eqref{H in III} with mixed singlet-triplet pairing. a) Zero temperature phase diagram, with fixed magnetic interaction, fixed triplet pairing interaction and varying intraband singlet pairing interaction. The $C_4$ symmetry breaking ultranodal ground state exists within our model when the triplet pairing and the $s$- and $d$-wave singlet pairing are all nearly degenerate. (b) Free energy versus temperature plot of various solutions at $W_1^s=0.8$ eV, $W_1^d=0.9$ eV. The system goes from the $d$-wave state to the $s+id$ state, then through first order transition to the $C_2$ ultranodal state at this particular choice of the interaction strength. (c,d) Normal (black) and Bogoliubov (colored) Fermi surfaces of (c) the ground state $C_2$ ultranodal state; (d) the metastable $C_4$ symmetric ultranodal solution. (e,f) Angular dependence of the various order parameters of the ultranodal states corresponding to panel (c) and (d) respectively. For panel (f), the triplet gaps are of the complex $p_x+ip_y$ form, and only the magnitude of the order parameters are plotted. Parameters used: $\mu_1=\mu_2={-}10$ meV, $m_1={-}8$ eV$^{-1}$, $m_2={-}4$ eV$^{-1}$, $\Delta\theta_c=\frac{\pi}{10}$, $\lambda_1=2\lambda_2=\frac{2}{\sqrt{5}}$, $J=4.35$ eV, $V=0.78$ eV, $W_1^s=0.45W_2^s$, $W_1^d=0.45W_2^d$. For panel (b-f), $W_1^s=0.8$ eV, $W_1^d=0.9$ eV, which is marked by the red star in panel (a).}
    \label{Fig3}
\end{figure*}

Note that the $m_i$ are the band masses and $\tilde{m}$ represents magnetization. Eq.\eqref{self-consistency2 in II} can then be written as
\begin{align*}
    \frac{1}{V}=\bar{N}\int_0^{2\pi}\frac{d\theta_\mathbf{k}}{2\pi}\int_{-\Omega_{c}+h_{k_l\sigma}}^{\Omega_{c}-h_{k_u\sigma}} d\bar{\epsilon}_{\mathbf{k}\sigma}\omega_p^2(\theta_\mathbf{k})\frac{th(\bar{E}_{\mathbf{k}\sigma}, h_{\mathbf{k}\sigma})}{2\bar{E}_{\mathbf{k}\sigma}}
    \numberthis \label{gap eq integral in II}
\end{align*}
where $\bar{N}=(N_1+N_2)/2$ is the averaged DOS of the two parabolic bands. If we assume that the k-sum is cut off when $|\epsilon_{1\mathbf{k}}|>\Omega_c$ or $|\epsilon_{2\mathbf{k}}|>\Omega_c$, then the $\epsilon$-integral in Eq.\eqref{gap eq integral in II} has cutoff $\Omega_c-h_{k_{l,u}\sigma}$ depending on $\Tilde{m}_\mathbf{k}$ (See Fig.~\ref{Fig1}(b), ~\ref{Fig2}(a)), thus also depending on $\theta_{\mathbf{k}}$.

For fixed temperature, normal state dispersion and $\Delta_{\sigma\sigma}$, the RHS of Eq.\eqref{gap eq integral in II} only depends on the magnetization on each of the bands, and a larger RHS leads to a smaller pairing strength $V$. Therefore we may say that a certain configuration of the magnetization-shifted normal state dispersion helps pairing, if it makes the RHS of Eq.\eqref{gap eq integral in II} larger than the value without magnetization. The magnetization $\tilde{m}_\mathbf{k}$ affects the RHS of Eq.\eqref{gap eq integral in II} by (i) changing the splitting $h_{\mathbf{k}\sigma}$ between the two subbands with the same spin if $\lambda_1\neq\lambda_2$; and (ii) shifting $\bar{\epsilon}_{\mathbf{k}\sigma}$, which in turn shifts the energy cutoffs $\Omega'_c$ as well as $h_{\mathbf{k}\sigma}$.
Note that when we say $h_{\mathbf{k}\sigma}$ changes, we mean when considering it as function of $\mathbf{k}$ rather than as a function of $\bar{\epsilon}_{\mathbf{k}\sigma}$. Examining Eq.~\eqref{h_k,sigma1}, it is obvious that  $\lambda_1\neq\lambda_2$ is the correct condition for $\tilde{m}_\mathbf{k}$ being able to alter $h_{\mathbf{k}\sigma}$ and not the condition $m_1\lambda_1\neq m_2\lambda_2$ as one might imagine by looking at Eq.\eqref{h_k,sigma}.

There are two general conditions that can help the interband equal spin pairing of spin $\sigma$. First, because $th(E,h)$ decreases monotonically as $|h|$ increases, smaller splitting $|h_{\mathbf{k}\sigma}|$  helps pairing, especially at places where the denominator $\bar{\epsilon}_{\mathbf{k}\sigma}$ is also small. This corresponds to shifting the two subbands' crossing point towards the Fermi level for $m_1\neq m_2$, or shifting the two subbands closer to each other for $m_1=m_2$. Secondly, a larger energy interval $[-\Omega_{c}+h_{k_l\sigma}, \Omega_{c}-h_{k_u\sigma}]$ where pairing is allowed also clearly helps pairing.

Guided by the above intuition, we discuss the following four cases: (i) $m_1=m_2$, $\lambda_1=\lambda_2=1/\sqrt{2}$. This corresponds to Fig.~\ref{Fig1}(a). The single band triplet pairing problem in 2D with fluctuating magnetism can be viewed as a special case of the interband pairing problem falling into this case, with $h_{\mathbf{k}\sigma}=0$ everywhere. Because $h_{\mathbf{k}\sigma}$ is the same for both spins regardless of the value of $\tilde{m}_\mathbf{k}$, both the integrand and the limits of the integral in Eq.\eqref{gap eq integral in II} are the same for both spins, there can be no non-unitary pairing in 2D. (ii) $m_1=m_2$, $\lambda_1\neq\lambda_2$. This corresponds to Fig.~\ref{Fig1}(b). Because $h_{\mathbf{k}\sigma}$ is reduced for one spin component and enlarged for the other spin, and consequently the energy interval for pairing is also enlarged for the favored spin, non-unitary solutions exist and { are} energetically favorable, even for pairing strength $V$ less than the unitary critical value $V_c$. This is shown in the phase diagram Fig.~\ref{Fig1}(c), where the unitary critical pairing strength is $V_c\approx0.42$ eV and the non-unitary phase persists down to $V\approx0.35$ eV when $\lambda_1$ is very different from $\lambda_2$. In panel \ref{Fig1}(d) we see that the non-unitary state where both $\Delta_{\uparrow\uparrow}$ and $\Tilde{m}$ are non-zero indeed arises spontaneously from the normal state where both order parameters are zero through a first order transition. In addition, the transition temperature for the nonunitary phase can be higher than the unitary $T_c$. (iii) $m_1\neq m_2$, $\lambda_1=\lambda_2=1/\sqrt{2}$. There are furthermore two different scenarios in this case. The first scenario is that the two parabolic bands cross each other already when there is no magnetization. This corresponds to Fig.~\ref{Fig2}. Because the crossing of the two subbands can be shifted closer to $E_F$ for one spin and away from $E_F$ for the other spin by any finite $\tilde{m}_\mathbf{k}$, a spontaneous non-unitary  phase also exists for this case and is shown by the phase diagram Fig.~\ref{Fig2}(b). In the second scenario where the two parabolic bands do not cross, interband pairing, unitary or non-unitary, is very difficult due to the separation of the bands and requires huge pairing strength. Therefore this scenario is excluded from our search for non-unitary pairing. (iv) $m_1\neq m_2$, $\lambda_1\neq\lambda_2$. This is the most general case where spontaneous non-unitarity is possible due to the interband nature of the the triplet pair. We will construct our microscopic Hamiltonian hosting BFS in the subsequent sections based on the Hamiltonian \eqref{H in II} with $m_1\neq m_2$ and $\lambda_1\neq\lambda_2$, and the BFS is largely a consequence of having singlet superconductivity coexisting with this nonunitary triplet order.

\section{2-band model with mixed singlet-triplet pair}
Now we add spin singlet pairing to the Hamiltonian in the previous section. To allow for the possibility of spontaneous $C_4$ symmetry breaking, we consider intraband pairing in both $s$- and $d$-wave channel, namely 
\begin{align*}
    H'=&H_0 +H_m+H_T\\
    &-\sum_{\substack{\mathbf{k},\mathbf{k'},i}}\big(W_i^s\omega_{s}(\theta_\mathbf{k})\omega_{s}(\theta_\mathbf{k'})+W_i^d\omega_{d}(\theta_\mathbf{k})\omega_{d}(\theta_\mathbf{k'})\big)\\
    &\ \ \ \ \ \ \ \ \ \ \ \ \ \times c^\dagger_{i\mathbf{k}\uparrow}c^\dagger_{i\mathbf{-k}\downarrow}c_{i\mathbf{-k'}\downarrow}c_{i\mathbf{k'}\uparrow}
    \numberthis \label{H in III}
\end{align*}

We use the $s$-wave form factor $\omega_{s}(\theta_\mathbf{k})=\cos^4 2\theta_\mathbf{k}$, which has accidental nodes along the $45^\circ$ directions, and the $d$-wave form factor $\omega_{d}(\theta_\mathbf{k})=\cos 2\theta_\mathbf{k}\omega_{s}(\theta_\mathbf{k})$. The $\theta_\mathbf{k}$-dependent gap function is $\Delta_{i}(\theta_\mathbf{k})=\omega_{s}(\theta_\mathbf{k})\Delta_{i}^s+\omega_{d}(\theta_\mathbf{k})\Delta_{i}^d$ and the singlet s/d-wave order parameters are determined self-consistently by $\Delta^{s/d}_i=W^{s/d}_i\sum_\mathbf{k} \omega_{s/d}(\theta_\mathbf{k})\langle c_{i\mathbf{-k}\downarrow}c_{i\mathbf{k}\uparrow}\rangle$. We note that the form factors assumed here respect the tetragonal symmetry, but are not of the lowest order in $\theta_{\mathbf{k}}$ that would be allowed. 
For now, this choice is made only because we did not find energetically favorable ultranodal solutions using lower order hamonics.
We postpone the discussion of the physical justifications of the singlet pairing form factors to the next section, where a more realistic 4-band model is considered.
The model is now solved within the mean field approximation numerically. For a particular choice of the interaction strength (see the caption of Fig.~\ref{Fig3}), we found, among other solutions to the self-consistency equations, a solution with non-unitary triplet $p_x$ pairing and $s+d$ singlet pairing, which has a twofold symmetric Bogoliubov Fermi surface.  We refer to this state as  the $C_2$ ultranodal state, and note that in the current model it exists only over a narrow range of parameters in the region of the phase diagram where singlet states and unitary triplet states are competitive. Fig.~\ref{Fig3}(b) compares the free energy of various solutions. At zero temperature, the $C_2$ ultranodal solution has the lowest free energy, therefore it is the ground state. For the same interaction strength at higher temperature, the system has a stable singlet $d$-wave pairing state and subsequently an $s+id$ state in a narrow temperature range. Fig.~\ref{Fig3}(c) and (e) show the Bogoliubov Fermi surface and the angular dependence of the order parameters of the $C_2$ ultranodal ground state. Clearly it is energetically favorable to have the singlet and the non-unitary triplet pair living on different parts of the Fermi surface, thus avoiding competition. In contrast, for the metastable $C_4$ symmetric solution shown in panel (d) and (f), the singlet and triplet pair cannot avoid each other (unless one of them is zero) so there is loss of total condensation energy.

\section{4-band model}
\begin{figure}
    \centering
    \includegraphics[width=0.95\linewidth]{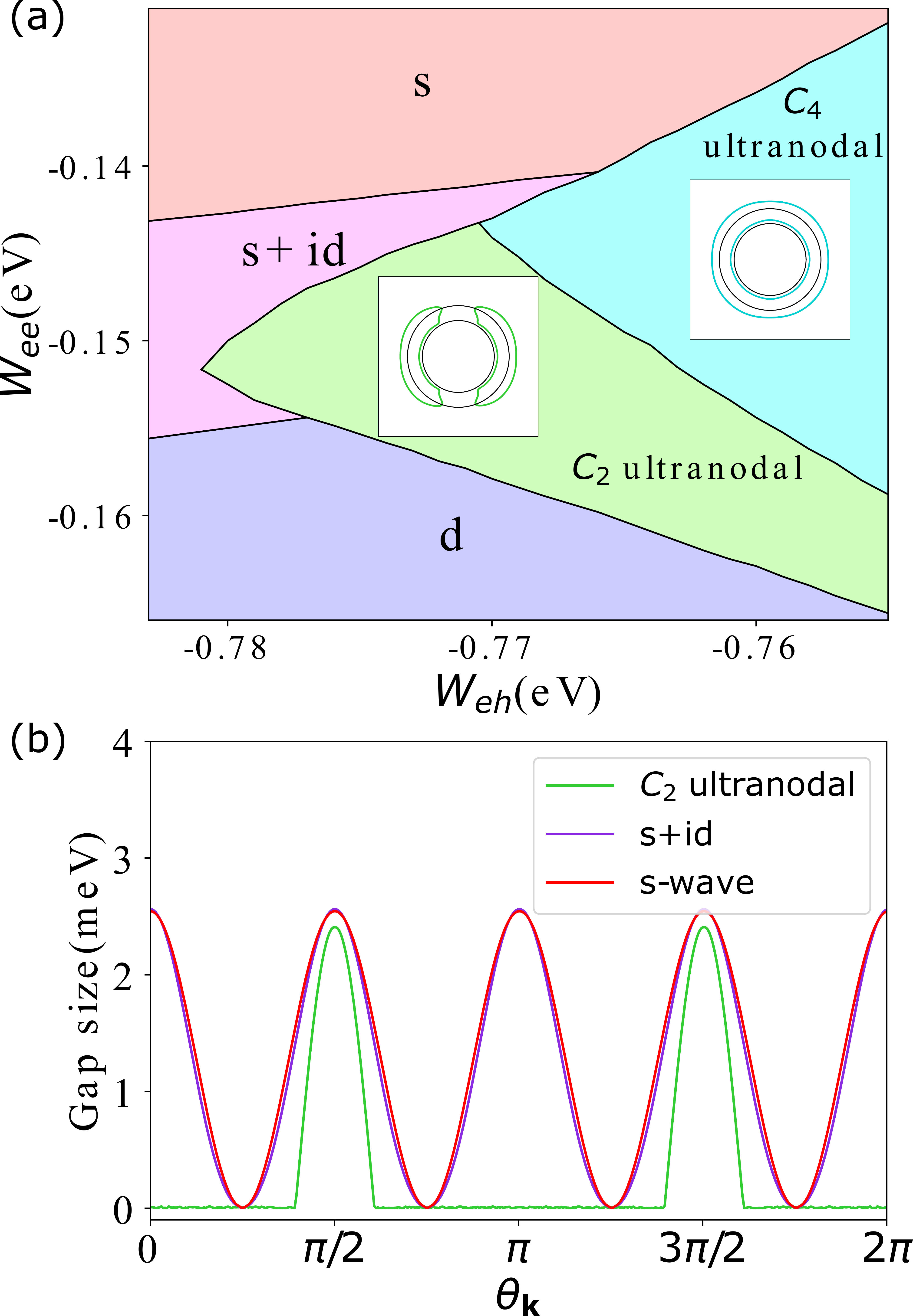}
    \caption{Results for the four band model Hamiltonian \eqref{H in IV}. Only the hole pockets near $\Gamma$ is shown and the $X$ and $Y$ pockets are not shown because they are assumed to be only trivially involved in the $s_\pm$ singlet pairing but not in other physical processes that are responsible for the Bogoliubov Fermi surfaces. (a) Phase diagram near zero temperature. Insets: typical shapes of the Bogoliubov Fermi surfaces of the $C_2$ and $C_4$ ultranodal ground states near the hole pockets. Parameters used: $\mu_h={-}10 $ meV, $ \mu_e=25 $ meV, $ m_1={-}8 $ eV$^{-1}$, $ m_2={-}4 $ eV$^{-1}={-}m_e$, $ \Delta\theta_c=\frac{\pi}{10}$, $\lambda_1=2\lambda_2=\frac{2}{\sqrt{5}}$, $a=0$, $b=0.2$, $J = 4.325$ eV, $ V = 0.775$ eV, $\frac{W_{eh}}{W'_{eh}}=0.45$. (b) The lowest quasiparticle excitation energy at all $\mathbf{k}$ along the angle $\theta_\mathbf{k}$, for typical $s$-wave, $s+id$ and $C_2$ ultranodal ground states in the phase diagram (panel(a)). For the $C_2$ ultranodal state, we used $W_{eh}=-0.775$ eV, $W_{ee}=-0.148$ eV.
    }
    \label{Fig4}
\end{figure}
We now consider a model which is crudely relevant for the tetragonal phase of FeSe$_{1-x}$S$_x$ ($x>0.17$) of interest,
consisting of four bands with 2D parabolic dispersions $\epsilon_{1,2}=\frac{k_x^2+k_y^2}{2m_{1,2}}-\mu_h,\ \epsilon_{3}=\frac{(k_x-\pi)^2+k_y^2}{2m_{e}}-\mu_e,\ \epsilon_{4}=\frac{k_x^2+(k_y-\pi)^2}{2m_{e}}-\mu_e$. The masses $m_1, m_2$ and the Fermi energy $\mu_{h}$ are taken to be negative, and $m_e$, $\mu_e$ to be positive, therefore bands 1 and 2 are hole bands degenerate at $\Gamma$, and bands 3 and 4 are electron bands centered at $X$ and $Y$ respectively.
We note that in the real material, there are some qualitative differences to our model such as the splitting due to spin orbit coupling at the $\Gamma$ point. Furthermore, the dispersion in the third dimension as detected experimentally leads to only one hole band crossing the Fermi level for some $k_z$\cite{Reiss2017,Coldea2021}; our model can be seen as an effective (average) model in two dimensions.
The Hamiltonian is
\begin{align*}
    H=&\sum_{\substack{\mathbf{k},\sigma\\i=\{1,2,3,4\}}}\epsilon_{i\mathbf{k}} c^\dagger_{i\mathbf{k}\sigma}c_{i\mathbf{k}\sigma} +H_m +H_T\\
    &+\sum_{\substack{\mathbf{k},\mathbf{k'}\\i,j=\{1,2,3,4\}, i\leq j}}\Gamma_{ij}(\mathbf{k}, \mathbf{k'})c^\dagger_{i\mathbf{k}\uparrow}c^\dagger_{i\mathbf{-k}\downarrow}c_{j\mathbf{-k'}\downarrow}c_{j\mathbf{k'}\uparrow}
    \numberthis \label{H in IV}
\end{align*}
The ferromagnetic term $H_m$ and the triplet interaction $H_T$ are the same as in Eq.\eqref{H in II} and involve only the two hole pocket at $\Gamma$ point. The last term is the repulsive $s_{\pm}$ pairing interaction. The effective pairing interaction in band space $\Gamma_{ij}(\mathbf{k},\mathbf{k'})$ should in principle depend on the orbital content of each band if deduced from the spin fluctuation theory\cite{Graser_2009}. Here we make a simplified assumption of an effective pairing interaction, namely
\begin{align}
    &\Gamma_{13}(\mathbf{k},\mathbf{k'})=\frac{W_{eh}}{W'_{eh}}\Gamma_{23}(\mathbf{k},\mathbf{k'})=W_{eh}\alpha_x(\theta_\mathbf{k})\\
    &\Gamma_{14}(\mathbf{k},\mathbf{k'})=\frac{W_{eh}}{W'_{eh}}\Gamma_{24}(\mathbf{k},\mathbf{k'})=W_{eh}\alpha_y(\theta_\mathbf{k})\\
    &\Gamma_{34}(\mathbf{k},\mathbf{k'})=W_{ee}\\
    &\Gamma_{12}(\mathbf{k},\mathbf{k'})=\Gamma_{ii}(\mathbf{k},\mathbf{k'})=0
\end{align}
The form factors on the hole bands $\alpha_x(\theta_\mathbf{k})$ and $\alpha_y(\theta_\mathbf{k})$ are related by $\alpha_x(\theta_\mathbf{k})=\alpha_y(\theta_\mathbf{k}+\frac{\pi}{2})$ due to the $C_4$ symmetry of the pairing interaction. For simplicity, the pairing interactions on the electron pockets are taken to be isotropic, i.e. $\Gamma_{ij}(\mathbf{k},\mathbf{k'})$ does not actually depend on $\mathbf{k'}$ when $j=3,4$.
As a consequence the gaps on the two electron pockets are isotropic and they have the same sign if the $(\pi,0), (0,\pi)$ repulsive pairing interaction dominates the $(\pi,\pi)$ one, i.e. if $W_{eh}\gg W_{ee}$. In this case, the gaps on the hole pockets acquire the $s$-wave form factor $\alpha_x(\theta_\mathbf{k})+\alpha_y(\theta_\mathbf{k})$. In the opposite limit if $W_{eh}\ll W_{ee}$, the electron pockets gaps acquire opposite signs and the hole pockets gaps have the $d$-wave form $\alpha_x(\theta_\mathbf{k})-\alpha_y(\theta_\mathbf{k})$. We assume that the $s$-wave form factor is $\alpha_x(\theta_\mathbf{k})+\alpha_y(\theta_\mathbf{k})\equiv \cos^2(2\theta_\mathbf{k})+a$. In accordance to this, we further make the assumption that
\begin{align*}
    \alpha_x(\theta_\mathbf{k}) \equiv (\cos^2(2\theta_\mathbf{k})+a)(\cos^2(\theta_\mathbf{k})+b)/(1+2b) \numberthis \label{form factor}
\end{align*}
where the latter constant $b$ tunes the overlap between $\alpha_x$ and $\alpha_y$ while keeping the $s$-wave form unchanged.  
We note that the $s$-wave form factor we use is indeed the lowest order harmonics suitable for $C_4$ symmetric gap functions with accidental minimal/nodes, while the $d$-wave form factor $\alpha_x-\alpha_y$ is not of the lowest order form, since the latter would be $\alpha_x-\alpha_y=\cos(2\theta_\mathbf{k})$. However, the particular choice \eqref{form factor} is made in order to mimic a situation where the $s$-wave and $d$-wave forms are similar in the shapes of the absolute values but only different in signs. This is likely to occur in multiband systems with orbital structure, when competing instabilities and nesting is present\cite{Romer2021,Graser_2009}.

Within mean field theory, this 4-band model \eqref{H in IV} can have two ultranodal phases at low temperature, as shown in Fig.\ref{Fig4}(a). First there is an ultranodal phase that spontaneously breaks the $C_4$ symmetry of the pairing interaction. The gap structure and the Bogoliubov Fermi surface is similar to what we have for the 2-band model in Sec. III., namely the singlet gap acquires the real $s+d$ form and the non-unitary triplet acquires the $p_x$ form, in order to avoid competition between $s+id$ and $p+ip$. In Fig.\ref{Fig4}(b) we show the calculated gap size as a function of angle, which is defined as the lowest quasiparticle excitation energy at all $\mathbf{k}$ along the angle $\theta_\mathbf{k}$. As expected, in a wide angle range, the quasiparticle excitation energy is zero for the $C_2$ ultranodal ground state. This result and the shape of the Bogoliubov Fermi surface is in agreement with the recent ARPES measurement\cite{nagashima2022discovery}. In addition, there is an ultranodal phase at low temperature that preserves the $C_4$ symmetry. We note that the corresponding BFS, as shown in the inset of Fig.\ref{Fig4}(a), has different shape from that of the metastable $C_4$ ultranodal solution of the two-band model.
\begin{figure}
    \centering
    \includegraphics[width=\linewidth]{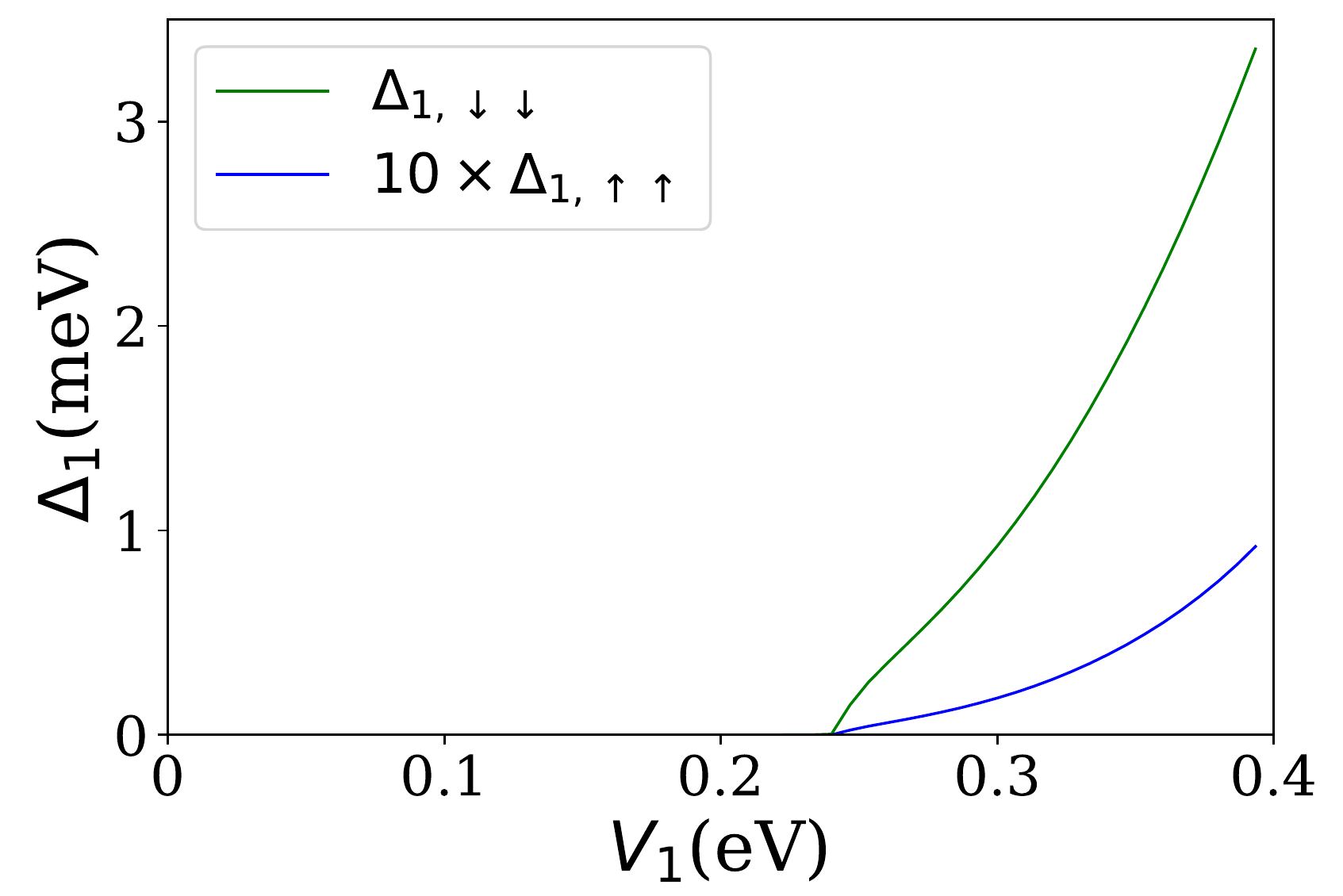}
    \caption{The induced intraband spin-triplet gap as a function of the perturbing intraband spin-triplet pairing strength $V_1$ on the $C_2$ ultranodal state. The unperturbed $C_2$ ultranodal state is obtained using the same set of parameter as in the caption of Fig.\ref{Fig4}. At $V_1=0.24$ eV$=0.31 V$ ($ V = 0.775$ eV) the intraband triplet gaps $\Delta_{1\sigma\sigma}$ become finite and gap out the BFS.}
    \label{Fig5}
\end{figure}

\section{discussion}
We have { exhibited} models with weak magnetic  and mixed singlet-triplet pairing interactions that can host $C_2$ ultranodal states as ground states. As temperature increases, our models undergo first order transitions from the ultranodal ground state to either singlet or triplet superconducting phases (see e.g. Fig.\ref{Fig3}(b)), and the free energies of the ultranodal solutions at these temperature are no longer the lowest. { There is, to our knowledge no experimental evidence of the tetragonal Fe(Se,S) system having multiple transitions below the superconducting $T_c$.   On the other hand,  current $T$-dependent evidence for the BFS is not sufficient to rule out such transitions.   In addition, we note that}   nematic fluctuations are significant even in the tetragonal phase above the critical S-substitution\cite{hosoi2016nematic}, and it has been pointed out\cite{chen2020nematicity} that in the presence of comparable $s$- and $d$- channel pairing interaction, it is energetically favorable for the nematic fluctuations to be stabilized along with an $s+d$ superconducting gap. The $s+d$ singlet gap structure is consistent with our $C_2$ ultranodal state. Therefore, we expect that if the { nematic field is accounted for and  allowed to order}, in addition to our proposed Hamiltonian Eq.\eqref{H in III} or \eqref{H in IV}, the free energy of our $C_2$ ultranodal state will be further lowered; {  and it is possible that this ultranodal phase becomes considerably more robust, expanding both in interaction parameter space and  to  higher temperatures.}

Another issue we would like to briefly address is the lack of inversion symmetry of our mixed-singlet-triplet mean field Hamiltonian. Although the Hamlitonian \eqref{H in III} and \eqref{H in IV} is symmetric under parity, unlike the mean field model we proposed in Refs. \cite{Setty2019,Setty2020}, which contains even parity interband spin-triplet pairing terms, the mean field Hamiltonian corresponding to our microscopic model in this paper (See Eq.\eqref{MFhamiltonian}) is not invariant under parity, because of the presence of both the even-parity intraband spin-singlet pair and odd-parity interband spin-triplet pair. Nevertheless, the existence of the Bogoliubov Fermi surface can be still understood as a consequence of having sign changes of the Pfaffian Pf$(\tilde{H_\mathbf{k}})$ across the Brillouin zone. In this case, despite the lack of inversion symmetry $U_PH_\mathbf{-k}U_P^\dagger\neq H_\mathbf{k}$ with $U_P=\mathbbm{1}$, there is an accidental symmetry $U_QH_\mathbf{-k}U_Q^\dagger= H_\mathbf{k}$ with $U_Q=\tau_z$. Here $\tau_z$ is the Pauli matrix in band space where the two hole bands are considered, and the electron bands are unchanged under this accidental symmetry. Using the unitary operator $U_Q$ instead of $U_P$, one can transform $H_\mathbf{k}$ into an anti-symmetric form and define the Pfaffian following the same line of thought as in Ref. \cite{Timm2017}.

As we just mentioned, the inversion symmetry of the microscopic Hamiltonian \eqref{H in IV} does not guarantee the inversion symmetry of the corresponding mean field Hamiltonian at low temperature. 
{ When the superconducting state spontaneously breaks inversion symmetry, the Bogoliubov Fermi surface can still exist, as was shown in the previous paragraph, or it can be gapped out. To see the latter point,} let's consider adding an intraband triplet paring interaction on the hole pockets to the Hamiltonian \eqref{H in IV},
\begin{align*}
    H_{T,i}=-V_i\sum_{\mathbf{k},\mathbf{k'},\sigma}\cos(\theta_\mathbf{k}-\theta_\mathbf{k'})c^\dagger_{i\mathbf{k}\sigma}c^\dagger_{i-\mathbf{k}\sigma}c_{i-\mathbf{k'}\sigma}c_{i\mathbf{k'}\sigma},\\ i=1,2
    \numberthis \label{intraband triplet}
\end{align*}
which still preserves the inversion symmetry { of \eqref{H in IV}. If a finite intraband triplet odd-parity pair $\Delta_{1,\downarrow\downarrow}$ is induced, the BFS we showed in Fig.\ref{Fig4} will be gapped out.} In Fig.\ref{Fig5} we see that for $V_1<{0.31}V$ there is no intraband triplet condensate induced by the additional term \eqref{intraband triplet} and the $C_2$ ultranodal solution remains the ground state. For $V_1>{0.31}V$ the solution acquires a finite $\Delta_{1,\downarrow\downarrow}$, which will gap out the BFS. The ultranodal state is stable against small intraband triplet pairing interaction, but will become fully gapped when the intraband triplet pairing interaction exceeds a critical value within our model. In materials with multiple orbital degrees of freedom where singlet and triplet pairing are assumed to arise from a spin-fluctuation mechanism, it is natural that interband pairing (with finite momentum transfer pair hopping) can be large if the leading instability is a singlet state. In this case, the triplet pairing interaction also has the same properties, i.e. large pairing interaction for large momentum transfer (interband) and small pairing interaction for small momentum transfer (intraband)\cite{Kreisel2022}. In this scenario, the coefficient $V_i$ obtained by  projecting onto the leading harmonic in Eq.~(\ref{intraband triplet}) is expected to be small. Whether the interband triplet pairing coefficient $V$ in Eq.~(\ref{H in II}) exceeds $V_i$ by a parametric  factor depends on details of the model such as shape of the Fermi surface and orbital weights\cite{Graser_2009,Kemper2010}.
 
In this work, we have assumed a ferromagnetic interaction in our microscopic models, which helps stabilize the non-unitary triplet pair and the BFS. However, there exist no strong signatures of ferromagnetic correlations in Fe based superconductors, although there are a few exceptions \cite{Wright2013,McLaughlin2021}.
{Previous theoretical studies have found 
stable spontaneous TRSB ultranodal states in the strong spin orbit coupling and high angular momentum $j=3/2$ scenario\cite{Menke2019}, where the non-unitarity has a different origin than the complex d-vector spin triplet pairing in our spin-1/2 scenario.} Therefore, we anticipate that spin-orbit coupling might be able to play the role of ferromagnetic interactions in our current model and stabilize an ultranodal state with non-unitary {spin triplet} pairing, but the actual form of  minimal spin-orbit terms that can stabilize such states in spin-1/2 models  requires further investigation.

{ Finally, we note that in our model, we have focused only on the case where the ground state is not an eigenstate of parity (case (d) in Ref. \cite{Setty2020} with a momentum dependent triplet pair). However, Bogoliubov Fermi surfaces are also possible in other situations. For example, they can occur when the pairing state is odd under both charge conjugation and inversion symmetries but preserves their product (case (b) in Ref. \cite{Setty2020}), or (case (c)) where the order parameter has a purely imaginary component. In the context of noncentrosymmetric superconductors, case (b) was discussed in Ref.  \cite{Link2020}.{  Recently, we learned that a similar scenario for the phenomenology of tetragonal FeSe,S  has been investigated by  Wu,  Amin, Yu, and Agterberg\cite{Wu_private}.} 
}
\section{Conclusions}

In this paper, we have studied Hamiltonians with triplet pairing  and itinerant ferromagnetic interactions, together  with a singlet pairing interaction, in a multiband scenario. We showed that such models, treated in self-consistent mean field theory, possess a well-defined Pfaffian and are equivalent to phenomenological Hamiltonians    expected to host surfaces of zero energy excitations in the superconducting state (ultranodal state).  Such  models  have been introduced in the context of the Fe-based superconductor Fe(Se,S) in the tetragonal phase, which seems to exhibit a residual density of states below $T_c$ without evidence of significant disorder. Within this framework, we have shown that  interband non-unitary triplet states can be stabilized by ferromagnetic fluctuations. In addition, when singlet pair order coexists with triplet order, the competition of various instabilities can lead to energetically favorable ultranodal ground states.

The self-consistent theory presented here allows a calculation of the temperature dependences of the various gaps in the model, and, in principle, a quantitative description of the topological transition to the ultranodal state.  As such, it is a significant step beyond the predictions of Ref. \cite{Setty2019}, and a good starting point to try to understand the properties of the Fe(Se,S) system. 

Depending on details, we find the system thus described may condense in an ultranodal state that may or may not preserve the underlying $C_4$ symmetry of the    crystal lattice in the tetragonal normal state. In the latter case the ground state of the model is only $C_2$ symmetric.    At present, the theory neglects orbital degrees of freedom, a discussion of which we postpone to a further study. However,  we anticipate that allowing for spontaneous orbital order or other electronic nematic orders will generally enhance the robustness of the $C_2$-symmetric ultranodal phase.

These theoretical findings may have connections to (nearly) ferromagnetic superconductors where non-unitary states could emerge without ferromagnetic order in the normal state, and to the Fe(Se,S) system where various experimental observations could be explained by the presence of a $C_2$ symmetric BFS.  We note that recently weak time reversal symmetry breaking in the Fe(Se,S) system has been detected in $\mu$SR experiments\cite{Matsuura_TRSB}, but at present there is no evidence of which we are aware of significant ferromagnetic correlations in the Fe(Se,S) system. However, ferromagnetic correlations have occasionally been reported in other iron-based systems\cite{Wright2013,McLaughlin2021}, and our analysis should serve as a motivation to further experimental searches  in this direction.

\section{Acknowledgements} The authors acknowledge useful discussions with D. Agterberg, A. Coldea and A. Nevidomskyy.  L.~F. acknowledges support by the European Union's Horizon 2020 research and innovation programme through the Marie Sk\l{}odowska-Curie grant SuperCoop (Grant No 838526). P.J.H. and Y.C. were  supported by  DOE grant number DE-FG02-05ER46236.  A.K. acknowledges support by the Danish National Committee for Research Infrastructure (NUFI) through the ESS-Lighthouse Q-MAT.\\ \newline
\noindent

\bibliography{microBFS.bib}

\appendix
\begin{widetext}
\section{Stability of ultranodal solutions in single band model}
{ In the work by N. I. Karchev, {\it et al.}\cite{karchev2001coexistence},
the problem} of singlet superconductivity coexisting with itinerant ferromagnetic order for a single band was studied. Starting from an interacting Hamiltonian with an exchange term of strength $J$ and a density-density interaction of strength $g$, they used a Hubbard-Stratonovich transformation to derive self-consistency equeations for the FM order parameter $M$ and SC order parameter $\Delta$. They showed that a solution to the self-consistency equations with both the FM and SC order parameters non-zero exists. This solution would correspond to a physical ultranodal state if it turns out to be stable. However, it was realized in Ref. \cite{shen2003breakdown} that this original solution in 3D does not minimize the free energy, thus is unstable. We show here in 2D the same conclusion holds, i.e. a solution with coexisting FM and SC order exists but the free energy is higher than the trivial singlet SC solution.

We begin with introducing the same mean field Hamiltonian and self-consistency equations as in Ref. \cite{karchev2001coexistence} Eq.(3-7), but instead of a 3D model let us consider a 2D one:
\begin{align}
    &H_{\mathrm{eff}}=\sum_{\mathbf{p}}\epsilon_p^{\uparrow} c_{\mathbf{p} \uparrow}^{\dagger} c_{\mathbf{p} \uparrow}+\epsilon_p^{\downarrow} c_{\mathbf{p} \downarrow}^{\dagger} c_{\mathbf{p} \downarrow}+\Delta^* c_{-\mathbf{p} \downarrow} c_{\mathbf{p} \uparrow}+ h.c.\\
    &\epsilon_p=\frac{p^2}{2 m^*}-\mu,\ \ \ \epsilon_p^{\uparrow}=\epsilon_p+\frac{J M}{2},\ \ \ \epsilon_p^{\downarrow}=\epsilon_p-\frac{J M}{2}\\
    &M=\frac{1}{2} \int \frac{d^2 p}{(2 \pi)^2}\left(1-n_p^\alpha-n_p^\beta\right) \numberthis \label{M in appendix}\\ &|\Delta|=\frac{|\Delta| g}{2} \int \frac{d^2 p}{(2 \pi)^2} \frac{n_p^\beta-n_p^\alpha}{\sqrt{\epsilon_p^2+|\Delta|^2}}
    \numberthis \label{self-consistency in appendix}
\end{align}
Here, $\epsilon_p$ is dispersion of free electrons and $n_p^{\alpha/\beta}$ are the distribution functions of the Bogoliubov Fermions with dispersion $E_p^{\alpha/\beta}=\frac{J M}{2}\pm\sqrt{\epsilon_p^2+|\Delta|^2}$.
We use the same notation as their Eq.(8) to denote the two momenta $p_F^\pm$ where the lower quasiparticle band
is at the Fermi level, $E_p^\beta=0$:
\begin{align}
    p_F^{\pm}=\sqrt{2 m^* \mu \pm m^* \sqrt{(J M)^2-4|\Delta|^2}}
\end{align}
In analog with Eq.(9) in Ref.~\cite{karchev2001coexistence}, for the 2D problem, Eq.\eqref{M in appendix} becomes
\begin{align*}
    M&=\frac{1}{8 \pi}\left[\left(p_F^{+}\right)^2-\left(p_F^{-}\right)^2\right]\\
     &=\frac{m^*}{4 \pi}\sqrt{(J M)^2-4|\Delta|^2}\numberthis
\end{align*}
This yields the same linear relation between $M$ and $|\Delta|$ as Eq.(12) in Ref.\cite{karchev2001coexistence} with a modified $r=\frac{Jm^*}{4\pi}$:
\begin{align}
    M=\frac{2}{J} \frac{r}{\sqrt{r^2-1}}|\Delta| \numberthis \label{M solution}
\end{align}
Note that to get the above relation we do not require $r\gg1$ as it was assumed in Ref. \cite{karchev2001coexistence} for 3D, but we still need $r>1$ for the solution to exist.\\

Eq.\eqref{self-consistency in appendix} is evaluated at zero temperature assuming an energy cutoff $\Lambda\gg |\Delta|$:
\begin{align*}
    1&=\frac{g}{4\pi}\left(\int_{\sqrt{2m^*(\mu-\Lambda)}}^{\sqrt{2m^*(\mu+\Lambda)}} d p \frac{p}{\sqrt{\epsilon_p^2+|\Delta|^2}}-\int_{p_F^{-}}^{p_F^{+}} d p \frac{p}{\sqrt{\epsilon_p^2+|\Delta|^2}}\right)\\
    &=\frac{gm^*}{8\pi}\left( \log\left(\frac{\sqrt{\Lambda^2+|\Delta|^2}+\Lambda}{\sqrt{\Lambda^2+|\Delta|^2}-\Lambda}\right)-\log\left(\frac{\frac{JM}{2}+\sqrt{(\frac{JM}{2})^2-|\Delta|^2}}{\frac{JM}{2}-\sqrt{(\frac{JM}{2})^2-|\Delta|^2}}\right)\right)\\
    &=\frac{gm^*}{8\pi}\left( \log\left(\frac{\sqrt{\Lambda^2+|\Delta|^2}+\Lambda}{\sqrt{\Lambda^2+|\Delta|^2}-\Lambda}\right)+\log\left(\frac{r-1}{r+1}\right)\right)\\
    &\approx\frac{gm^*}{4\pi}\left( \log\left(\frac{2\Lambda}{|\Delta|}\right)+\frac{1}{2}\log\left(\frac{r-1}{r+1}\right)\right)\numberthis \label{gap eq in Appendix}
\end{align*}
The last term on the RHS renormalizes the energy cutoff $\Lambda$, and therefore renormalizes $\Delta$, in the sense that 
\begin{align}
    \Delta=\sqrt{\frac{r-1}{r+1}}\Delta_0 \numberthis \label{Delta solution}
\end{align}
where we use $\Delta_0$ to denote the gap in the trivial singlet SC solution to Eq.\eqref{M in appendix}\eqref{self-consistency in appendix} with no FM order ($M=0$), equivalently, the solution to \eqref{gap eq in Appendix} without the last term on the RHS. Next we compute the condensation energy for the solution \eqref{M solution}\eqref{Delta solution} at $T=0$:
\begin{align*}
    &F(M,\Delta,T=0)-F(0,0,T=0)\\
    =&\frac{JM^2}{2}+\frac{|\Delta|^2}{g}-\int_{p_F^-}^{p_F^+}\frac{d^2p}{(2\pi)^2}\left|\epsilon_p\right|
    -\left(\int_{\sqrt{2m^*(\mu-\Lambda)}}^{\sqrt{2m^*(\mu+\Lambda)}}-\int_{p_F^-}^{p_F^+}\right)\frac{d^2p}{(2\pi)^2}\left(\sqrt{\epsilon_p^2+|\Delta|^2}-|\epsilon_p|\right)
    \numberthis \label{F in appendix}
\end{align*}
The last term is
\begin{align*}
    &\left(\int_{\sqrt{2m^*(\mu-\Lambda)}}^{\sqrt{2m^*(\mu+\Lambda)}}-\int_{p_F^-}^{p_F^+}\right)\frac{d^2p}{(2\pi)^2}\left(\sqrt{\epsilon_p^2+|\Delta|^2}-|\epsilon_p|\right)\\
    =&\frac{m^*}{\pi}\int_{\sqrt{(\frac{JM}{2})^2-|\Delta|^2}}^{\Lambda}d\epsilon\left(\sqrt{\epsilon^2+|\Delta|^2}-|\epsilon|\right)\\
    =&\frac{m^*}{2\pi}\left(\Lambda\left(\sqrt{\Lambda^2+|\Delta|^2}-\Lambda\right)-\frac{1}{r+1}|\Delta|^2+|\Delta|^2\log\left(\frac{\sqrt{\Lambda^2+|\Delta|^2}+\Lambda}{\sqrt{\frac{r+1}{r-1}}|\Delta|}\right) \right)\\
    =&\frac{m^*}{2\pi}\left(\frac{(r-1)^2}{2(r+1)^2}|\Delta_0|^2+\frac{r-1}{r+1}|\Delta_0|^2\log\left(\frac{2\Lambda}{|\Delta_0|}\right) \right)\numberthis
\end{align*}
To obtain the last line we used $\Lambda\gg|\Delta|$ and $\Delta=\sqrt{\frac{r-1}{r+1}}\Delta_0$. The other integral in Eq.\eqref{F in appendix} is
\begin{align}
    \int_{p_F^-}^{p_F^+}\frac{d^2p}{(2\pi)^2}\left|\epsilon_p\right|
    =\frac{m^*}{2\pi}\frac{1}{r^2-1}|\Delta|^2=\frac{m^*}{2\pi}\frac{1}{(r+1)^2}|\Delta_0|^2
\end{align}
And we also have the second term in Eq.\eqref{F in appendix}
\begin{align}
    \frac{|\Delta|^2}{g}=\frac{r-1}{r+1}\frac{|\Delta_0|^2}{g}=\frac{m^*}{2\pi}\frac{r-1}{r+1}|\Delta_0|^2\log\left(\frac{2\Lambda}{|\Delta_0|}\right)
\end{align}
We want to compare the condensation energy \eqref{F in appendix} with the condensation energy of the trivial SC-only solution:
\begin{align}
    F(0,\Delta_0,T=0)-F(0,0,T=0)=-\frac{m^*}{4\pi}\Delta_0^2
\end{align}
Putting everything together we get
\begin{align}
    F(M,\Delta,T=0)-F(0,\Delta_0,T=0)=\frac{JM^2}{2}+\frac{m^*}{2\pi}\frac{2r-1}{(r+1)^2}|\Delta_0|^2>0
\end{align}
This shows that the trivial SC-only solution is always energetically more favorable than the solution with both SC and FM order parameters nonzero in 2D.

\section{Details of the mean field treatment}
We show here explicitly the forms of the mean field Hamiltonian and free energy of Hamiltonian \eqref{H in IV} in the main text.
Applying a standard mean field treatment, Eq.\eqref{H in IV} becomes
\begin{align*}
    H=&\sum_{\substack{\mathbf{k},\sigma\\i=\{1,2,3,4\}}}\epsilon_{i\mathbf{k}} c^\dagger_{i\mathbf{k}\sigma}c_{i\mathbf{k}\sigma}
    -\sum_{\substack{\mathbf{k},\sigma\\i=\{1,2\}}}\sigma J\lambda_i\sum_{\mathbf{k'}}\hat{\Lambda}(\mathbf{k},\mathbf{k'})(\lambda_1m_{1\mathbf{k'}}+\lambda_2m_{2\mathbf{k'}})c^\dagger_{i\mathbf{k}\sigma}c_{i\mathbf{k}\sigma}\\
    &-\Bigg(\sum_{\mathbf{k},\sigma}(\cos{\theta_\mathbf{k}}\Delta_{px,\sigma\sigma}+\sin{\theta_\mathbf{k}}\Delta_{py,\sigma\sigma})c^\dagger_{1\mathbf{k}\sigma}c^\dagger_{2-\mathbf{k}\sigma}\\
    &\ \ \ \ +\sum_{\substack{\mathbf{k}\\i=\{1,2\}}}(\alpha_x(\theta_\mathbf{k})\Delta_{sx,i}+\alpha_y(\theta_\mathbf{k})\Delta_{sy,i})c^\dagger_{i\mathbf{k}\uparrow}c^\dagger_{i-\mathbf{k}\downarrow} +\sum_{\substack{\mathbf{k}\\i=\{3,4\}}}\Delta_{i}c^\dagger_{i\mathbf{k}\uparrow}c^\dagger_{i-\mathbf{k}\downarrow} +h.c.\Bigg)\\
    &+\frac{J}{2} \sum_{\mathbf{k},\mathbf{k'},i,j}\Lambda_{ij}(\mathbf{k},\mathbf{k'})m_{i\mathbf{k}}m_{j\mathbf{k'}}+\frac{|\Delta_{px\uparrow\uparrow}|^2+|\Delta_{px\downarrow\downarrow}|^2 + |\Delta_{py\uparrow\uparrow}|^2+|\Delta_{py\downarrow\downarrow}|^2}{V}\\
    &+\frac{\Delta_{sy1}\Delta_3^*+\Delta_{sx1}\Delta_4^*-\frac{W_{ee}}{W_{eh}}\Delta_{sx1}\Delta_{sy1}^*+h.c.}{W_{eh}}
    \numberthis \label{MFhamiltonian}
\end{align*}

The mean field order parameters are determined by the following self-consistency equations:
\begin{align}
    &m_{i\mathbf{k}}=\sum_{\sigma}\sigma\langle c^\dagger_{i\mathbf{k}\sigma}c_{i\mathbf{k}\sigma}\rangle,\\
    &\Delta_{px\uparrow\uparrow}=V\sum_\mathbf{k} \cos{\theta_\mathbf{k}}\langle c_{2-\mathbf{k}\uparrow}c_{1\mathbf{k}\uparrow}\rangle, \ \ \  \Delta_{py\downarrow\downarrow}=V\sum_\mathbf{k} \sin{\theta_\mathbf{k}}\langle c_{2-\mathbf{k}\downarrow}c_{1\mathbf{k}\downarrow}\rangle,\\
    &\Delta_{sx1}=W_{eh}\sum_\mathbf{k}\langle c_{3-\mathbf{k}\downarrow}c_{3\mathbf{k}\uparrow}\rangle, \ \ \ 
    \Delta_{sy1}=W_{eh}\sum_\mathbf{k}\langle c_{4-\mathbf{k}\downarrow}c_{4\mathbf{k}\uparrow}\rangle, \\
    &\Delta_{sx2}=\frac{W'_{eh}}{W_{eh}}\Delta_{sx1}, \ \ \Delta_{sy2}=\frac{W'_{eh}}{W_{eh}}\Delta_{sy1},\\
    &\Delta_3=\sum_\mathbf{k}\alpha_x(\theta_\mathbf{k})\left(W_{eh}\langle c_{1-\mathbf{k}\downarrow}c_{1\mathbf{k}\uparrow}\rangle+W'_{eh}\langle c_{2-\mathbf{k}\downarrow}c_{2\mathbf{k}\uparrow}\rangle\right)+W_{ee}\langle c_{4-\mathbf{k}\downarrow}c_{4\mathbf{k}\uparrow}\rangle,\\ &\Delta_4=\sum_\mathbf{k}\alpha_y(\theta_\mathbf{k})\left(W_{eh}\langle c_{1-\mathbf{k}\downarrow}c_{1\mathbf{k}\uparrow}\rangle+W'_{eh}\langle c_{2-\mathbf{k}\downarrow}c_{2\mathbf{k}\uparrow}\rangle\right)+W_{ee}\langle c_{3-\mathbf{k}\downarrow}c_{3\mathbf{k}\uparrow}\rangle
\end{align}
The self-consistency equations can be solved numerically at a given temperature. Once we get a solution to the self-consistency equations, the mean field free energy of the corresponding solution can be calculated using
\begin{align*}
    F=&-T\sum_{\mathbf{k},n}\ln{(1+e^{-\beta E_{n \mathbf{k}}})}-\frac{1}{2}\sum_{\mathbf{k},n} E_{n \mathbf{k}}+\sum_{\mathbf{k},i}\epsilon_{i\mathbf{k}}\\
    &+\frac{J}{2} \sum_{\mathbf{k},\mathbf{k'},i,j}\Lambda_{ij}(\mathbf{k},\mathbf{k'})m_{i\mathbf{k}}m_{j\mathbf{k'}}+\frac{|\Delta_{px\uparrow\uparrow}|^2+|\Delta_{px\downarrow\downarrow}|^2+ |\Delta_{py\uparrow\uparrow}|^2+|\Delta_{py\downarrow\downarrow}|^2}{V}\\
    &+\frac{\Delta_{sy1}\Delta_3^*+\Delta_{sx1}\Delta_4^*-\frac{W_{ee}}{W_{eh}}\Delta_{sx1}\Delta_{sy1}^*+h.c.}{W_{eh}}\numberthis
\end{align*}
Here $n$ is a quasiparticle band index and $E_{n \mathbf{k}}$ runs over all the \emph{positive} eigenvalues of the mean field Hamiltonian \eqref{MFhamiltonian}.

\end{widetext}
\end{document}